\documentclass{optica-article}

\journal{opticajournal} 

\articletype{Research Article}

\usepackage{bm}
\usepackage{makecell}

\begin{document}

\title{High-precision programming of large-scale ring resonator circuits with minimal pre-calibration}

\author{Shaojie Liu,\authormark{1,†} Tengji Xu,\authormark{1,†} Benshan Wang,\authormark{1} Dongliang Wang,\authormark{1} Qiarong Xiao,\authormark{1} and Chaoran Huang\authormark{1,*}}

\address{\authormark{1}Department of Electronic Engineering, The Chinese University of Hong Kong, Shatin, Hong Kong SAR, China}

\email{\authormark{*}crhuang@ee.cuhk.edu.hk} 


\begin{abstract*} 
Microring resonators (MRRs) are essential components in large-scale photonic integrated circuits (PICs), but programming these circuits with high precision and efficiency remains an unsolved challenge. Conventional methods rely on complex calibration processes that are both time-consuming and often inaccurate, limiting the scalability of PICs. This work introduces an innovative control method called chip-in-the-loop optimization (ChiL) that addresses this challenge by offering high scalability, precision, fast convergence, and robustness. ChiL reduces the calibration complexity for an $N$ devices system from $O(k^N)$ to a single-shot measurement, while maintaining a record-high precision over 9-bit in the presence of system imperfections, including fabrication variances, thermal crosstalk, and temperature drift. Using ChiL, we experimentally demonstrate a photonic solver for computing matrix eigenvalues and eigenvectors with errors on the order of  $10^{-4}$. Additionally, we achieve a photonic neural network (PNN) with accuracy and a confusion matrix identical to those of digital computers. ChiL offers a practical approach for programming large-scale PICs and bridges the gap between analog photonic and digital electronic computing and signal processing in both scale and precision.
\end{abstract*}

\section{Introduction}
\begin{figure}[ht!]
\centering\includegraphics[width=1\linewidth]{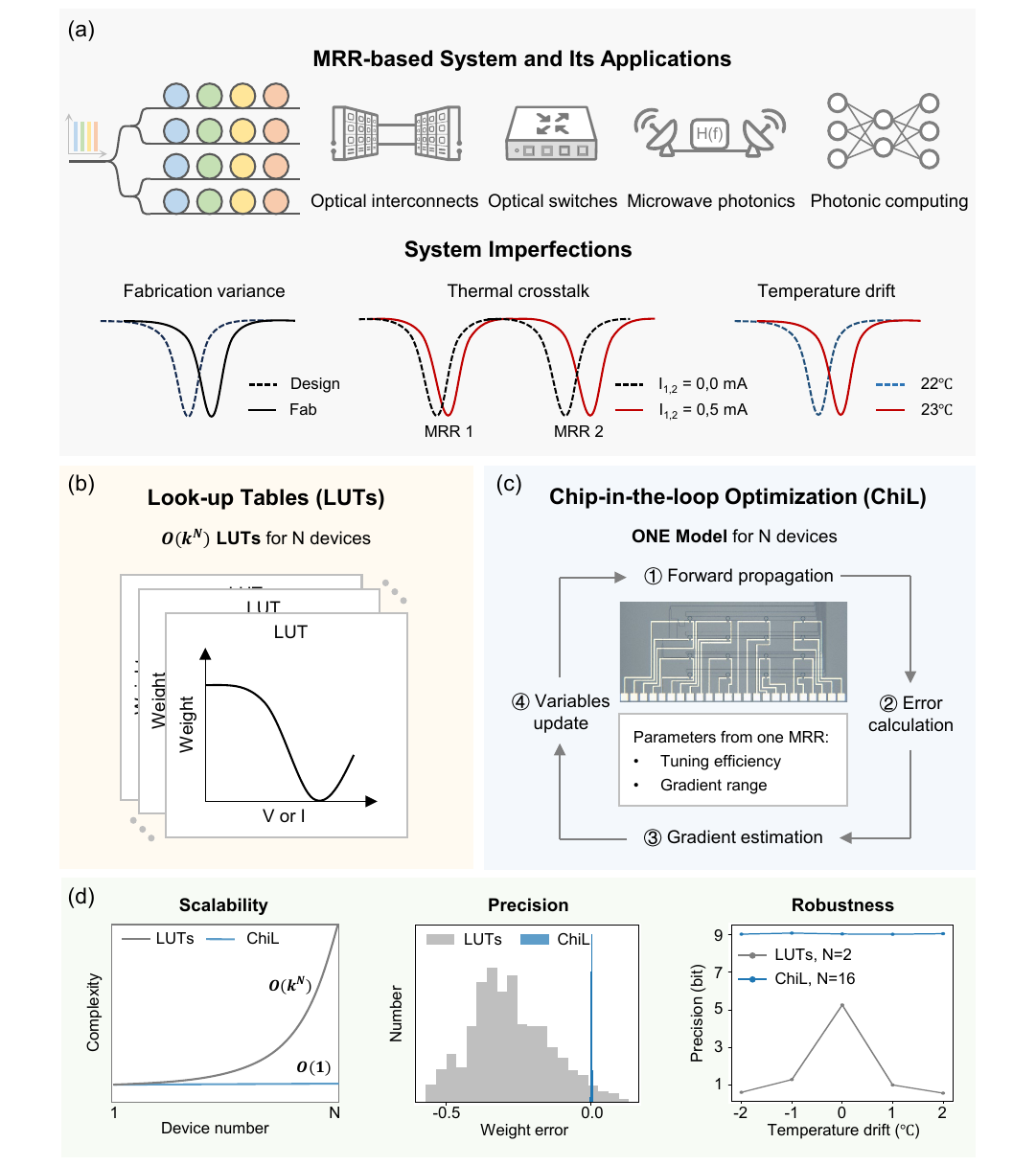}
\caption{
\textbf{Chip-in-the-loop optimization for MRR circuits} (a) MRR-based system with imperfections. (b) Previous methods of pre-calibration and building LUTs. (c) Concept of ChiL for MRR-based system. (d) Comparison of scalability, precision and robustness of LUTs and ChiL. LUTs: Look-up-tables, ChiL: Chip-in-the-loop optimization. 
}
\label{Fig1}
\end{figure}

The advancement of integrated photonic technology has enabled the fabrication of increasingly larger and more complex photonic systems on a chip. This trend toward large-scale photonic integrated circuits (PICs) holds the potential to become a transformative technology, addressing key challenges across various fields, including telecommunications\cite{xu_micrometre-scale_2005}, data centers\cite{wu_mode-division_2017,rizzo_massively_2023}, quantum technology\cite{wang_integrated_2020,pelucchi_potential_2022}, artificial intelligence\cite{shen_deep_2017,tait_neuromorphic_2017,feldmann_parallel_2021,huang_silicon_2021,zhou_photonic_2022,fu_optical_2024}, and 3D imaging\cite{sun_large-scale_2013}. In these large-scale PIC systems, integrating high-speed photonic analog processing systems with digital electronic control offers a promising avenue to overcome the limitations posed by the end of Moore’s law\cite{bogaerts_programmable_2020,perez-lopez_multipurpose_2020}. Microring resonators (MRRs) are versatile components widely employed in large-scale PICs, enabling a range of applications such as optical interconnects\cite{miller_rationale_2000}, optical switches\cite{poon_cascaded_2009,jayatilleka_photoconductive_2019}, microwave and optical signal filters\cite{capmany_tutorial_2006,liu_synthesis_2011,xu_high_2019}, pulse shaping\cite{cohen_silicon_2024}, photonic computing\cite{tait_neuromorphic_2017,huang_silicon_2021,zhang_system--chip_2024}, and others. MRRs provide several advantages, including a compact footprint, energy-efficient tuning and modulation, and high sensitivity\cite{bogaerts_silicon_2012}. However, its sensitivity, while advantageous in many ways, presents a double-edged sword. It intensifies the challenge of controlling MRRs in the presence of nonideal component behaviors, including fabrication-yield distributed errors\cite{chrostowski_impact_2014}, thermal crosstalk\cite{zhang_silicon_2022}, and various dynamic noises such as temperature drift, as shown in Fig. \ref{Fig1}(a).

Conventional approaches to programming MRR-based systems typically rely on constructing lookup tables (LUTs)\cite{huang_demonstration_2020,zhang_silicon_2022,cheng_self-calibrating_2023,liu_single-monitor_2024,bai_microcomb-based_2023} that define the relationship between the control variables (such as currents or voltages) and the states (typically transmission) of MRRs. Building such LUTs requires meticulous pre-calibration, not only for individual MRRs but also for the thermal crosstalk between them, as shown in Fig. \ref{Fig1}(b). As system complexity increases, the demands for calibration and programming grow exponentially due to thermal crosstalk, scaling with the component count at $O(k^N)$, where k is the number of points in one LUT, and N is the device number. This complexity can be overwhelming for large-scale PICs and severely limits the scalability and practicality of these systems.

In this work, we propose a highly scalable control method called chip-in-the-loop optimization (ChiL) to address this challenge. Our approach reduces the calibration complexity from $O(k^N$) for every MRR to a single-shot measurement on only one MRR, while achieving and maintaining record-high precision as the system scales up. Our method eliminates the need for precise calibration of every MRR and their thermal crosstalk. Instead, it only requires characterizing one MRR to extract key parameters, which are then applied to all MRRs. (Fig. \ref{Fig1}(c)). Our method is simple and efficient, while demonstrating the precision over 9-bit on a 4×4 MRR array, which is the highest precision achieved at this system scale. Importantly, system imperfections, including fabrication variances, thermal crosstalk, and temperature drift over ±2°C, do not degrade the precision of our method.

Enabling high-precision MRR programming is crucial for computing-related applications\cite{zhang_silicon_2022,cheng_self-calibrating_2023}. We demonstrate the importance of our method through two fundamental computing tasks. The first one is solving eigenvectors and eigenvalues of matrices, which has broad implications for data compression\cite{shlens_tutorial_2014}, computer vision\cite{draper_recognizing_2003}, solving differential equations\cite{gary_matrix_1970}, and so on. With ChiL, we solve matrices’ eigenvector and eigenvalue with errors on the order of $10^{-4}$, which is 1000 times smaller than previous work\cite{liao_matrix_2022}. The second application is in PNNs. We demonstrate 10-class handwritten digit recognition using the MRR-based PNN programmed by ChiL. Notably, our method not only achieves the same classification accuracy of 97.0\% as a digital computer but also produces a confusion matrix that is exact to that of the digital computer.

Our method leverages novel optimization techniques, providing a practical solution for programming large-scale PICs. It effectively bridges the gap in scale and precision between analog optical computing and digital electronic computing. Our method can be extended to various large-scale PICs, unlocking a wide range of applications, including MRR-based systems for microwave and optical filters, optical interconnects, optical communications, and more.

\section{Principle of Chip-in-th-loop Optimization}

MRRs are widely used in large-scale PICs for various applications. To illustrate the working principle of our method, we use the MRR weight bank for matrix-vector multiplication. Matrix-vector multiplication is performed with a "broadcast and weight" structure\cite{tait_broadcast_2014}, where the transmission of each MRR corresponds to a matrix element (or weight). The weight value is programmed via a microheater based on the thermo-optic effect. However, achieving precise programming of MRR weight banks is challenging due to both system imperfections, such as fabrication variance and thermal crosstalk, as well as dynamic noise sources like temperature drift. These issues can cause the MRR’s transmission to deviate from expected values, as shown in Fig. \ref{Fig1}(a).

To program such a system, common methods require developing a set of LUTs by calibrating the relationship between the current applied to each MRR and the resulting weight, as illustrated in Fig. \ref{Fig1}(b). Without considering thermal crosstalk, the total number of measurements needed to build the LUTs for a system with $N$ devices and $k$ points per LUT would be $kN$. However, programming MRRs based solely on these LUTs is inaccurate due to thermal crosstalk. When thermal crosstalk is considered, these MRRs become interdependent, meaning that the weight of an MRR depends not only on its own current but also on the currents applied to all other MRRs. Fully capturing these effects requires $k^N$ measurements, which increases exponentially with the device number and becomes unmanageable in large-scale systems. Moreover, any ambient changes, such as temperature drift, require recalibrating the whole system, further complicating the process.

The proposed ChiL method eliminates the need for tedious calibration and recalibration, enabling precise control of N MRRs simultaneously using only a single model. This model is simply a few parameters including thermal tuning efficiency and the maximum and minimum slope of the MRR’s transmission function, both of which can be obtained by sweeping current on an MRR and measuring its transmission spectrum (see Supplementary 1). These parameters are then shared when controlling all other MRRs, even though they are different due to fabrication variance. As a result, the total number of measurements remains constant and minimal regardless of device number. 

The concept of ChiL is illustrated in Fig. \ref{Fig1}(c). The key to accurately controlling all MRRs using a single model is incorporating the chip within the optimization loop. This integration allows monitoring weight at the chip output real-time, and subsequently reduce weight error using a gradient-based optimization method. In this way, even though the calibrated model does not dictate the exact currents to control all MRRs, it can still effectively indicate the correct direction for optimization through the estimated gradient and guide variables update throughout the process.

The optimization loop consists of four key steps: (1) forward propagation through the chip under operation, (2) error signal measurement from the chip output, (3) gradient estimation based on the error signal and the model, and (4) finally, control variables update based on the estimated gradient. The operation procedures are detailed as follows. Forward propagation can be expressed as $\bm{y}=\bm{W}(\bm{P})\bm{x}$, where $\bm{x}\in \mathbb{R}^n$ is the input vector, $\bm{y}\in \mathbb{R}^m$ is the output vector,$ \bm{W}\in \mathbb{R}^{m\times n}$ is the transmission matrix of the MRR weight bank, determined by the control variables $\bm{P} \in \mathbb{R}^{m\times n}$, where $m$ and $n$ denote the row and column number of the weight bank. In this system, $\bm{x}$ and $\bm{y}$ are signals encoded by the light intensity on different wavelengths, and $\bm{P}$ is normalized electrical powers applied to each MRR. Using an identity matrix $\bm{I}\in \mathbb{R}^{n\times n}$ as the input, we measure the weight of each MRR from the chip output based on $\bm{Y}=\bm{W}(\bm{P})\bm{I}=\bm{W}(\bm{P})$ (see Supplementary 1 for more details). Given a target weight matrix $\hat{W}$, the weight error can be defined as: $\bm{E}=\bm{W}(\bm{P})-\hat{\bm{W}}$. 

Gradient-based searching can efficiently determine the control variables $\bm{P}$ to minimize the error, especially in large-scale systems. However, the gradient of a physical system can only be estimated because it is non-differentiable\cite{wright_deep_2022}. Previous finite-difference methods\cite{shen_deep_2017,bandyopadhyay_single_2022,bai_microcomb-based_2023}  for gradient estimation requires additionally perturbating every control variables and measuring error in each iteration, which limits programming efficiency. Here we propose an efficient gradient estimation method that learns the gradient from iteration history. The gradient of weights with respect to the control variables is estimated by:
\begin{equation}
\partial\bm{W}/\partial\bm{P}=(\bm{W}_k-\bm{W}_{k-1})/(\bm{P}_k-\bm{P}_{k-1})
\label{eq1}
\end{equation}
where $k$ and $k-1$ indicates $\bm{W}$ and $\bm{P}$ in the current and previous iteration, separately, and $/$ denotes element-wise division. The estimated gradient may be unstable due to thermal crosstalk and dynamic noise, so we constrain the absolute value of it to a range $[G_{min},G_{max}]$, where $G_{min}$ and $G_{max}$ are minimum and maximum transmission slopes obtained from the calibrated model. Then we update all control variables simultaneously based on the rule given by:
\begin{equation}
\bm{P}_{k+1}=\bm{P}_{k}-\eta\frac{\bm{E}}{\partial\bm{W}/\partial\bm{P}}
\label{eq2}
\end{equation}
where $\eta$ is the learning rate, set as 0.5 in our experiment. Here we use Newton’s method, which converges faster than the conventional gradient descent method. Unlike conventional gradient descent, where the gradient $\partial\bm{W}/\partial\bm{P}$ is multiplied, we use it as a divisor. This strategy leads to a larger step size at the flat region and a smaller step size at the steep region of MRR’s transfer function, making the current weight converge to the target weight faster.

Our parameter update rule does not explicitly account for thermal crosstalk, but it is automatically compensated during our training process. This is because we update all control variables simultaneously in one iteration, rather than tuning each MRR to the target weight individually. By doing this, the errors caused by thermal crosstalk are detected in the current iteration  and guide the variable updates based on Eq. \ref{eq2}. Therefore ChiL is much simpler and more scalable compared to alternative solutions such as data-driven methods or building neural networks to model the system accurately\cite{wright_deep_2022,zheng_dual_2023}, which limits the scalability and adaptability to ambient fluctuations like temperature changes. The comparison of scalability, precision, and robustness of controlling MRR weight bank with LUTs and ChiL are shown in Fig. \ref{Fig1}(d). 

\section{Results}

\begin{figure}[ht!]
\centering\includegraphics[width=1\linewidth]{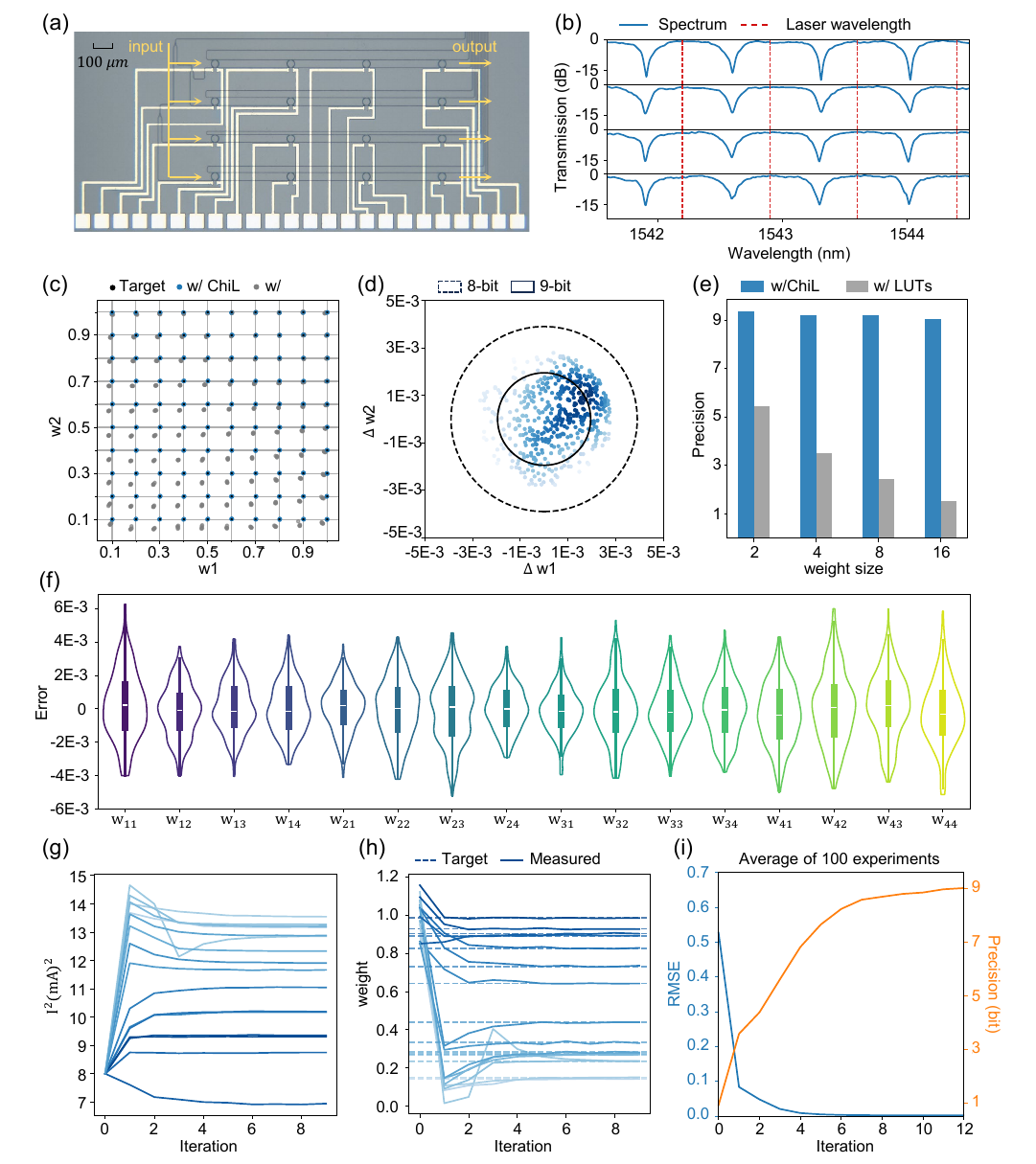}
\caption{
\textbf{MRR weight bank with over 9-bit precision} (a) A microscope photo of the 4×4 on-chip MRR weight bank. (b) Transmission spectra of through ports of MRR weight bank after initial alignment. (c) Target and measured weights of programming 2 MRRs with ChiL and LUTs. (d) Weight errors of programming 2 MRRs with ChiL. (e) Weight precision of programming increasing size of MRR weight bank with ChiL and LUTs. (f) Error distribution of programming a 4×4 MRR weight bank with ChiL. (g-h) Iteration history of control variables(g) and weights(h) in one experiment. i Iteration history of average weight’s RMSE and corresponding precision in 100 experiments. RMSE: root-mean-square-error.
}
\label{Fig2}
\end{figure}

\subsection{Scalable MRR circuit programming with over 9-bit precision}
We demonstrate the effectiveness of ChiL using a 4×4 silicon MRR weight bank chip as illustrated in Fig. \ref{Fig2}(a). Each MRR within the 1×4 weight bank has a slight difference in radius, leading to a unique resonant wavelength that aligns with the wavelength of the corresponding input laser. By tuning the resonant wavelength of each MRR using an embedded TiW heater., the power distribution of light between the Drop and Through ports can be precisely controlled, allowing for tunable weights. Details of the chip design and fabrication are provided in Supplementary 1.

To avoid spectral aliasing and align the resonant wavelengths within the same column, initial currents, determined by the tuning efficiency in the calibrated model, are applied to each MRR (see supplementary 1). The weight bank spectra after alignment, along with the working wavelengths of the four lasers, are shown in Fig. \ref{Fig2}(b). We use the square of current as the control variable, as it is nearly linear to the resonance shift. The input signals are modulated by four variable optical attenuators, and the outputs are detected by a 4-channel optical power meter. Further details of the experimental setup are provided in supplementary 1. 

To demonstrate the precision improvement of ChiL compared to traditional methods that rely on LUTs (see Supplementary 1), we start by programming two MRRs located at the first column of the weight bank.  To show ChiL's versatility across the entire feasible weight range and its repeatability under dynamic noise, we set 100 groups of target weights, evenly distributed between 0 and 1, and test each group for five times. Fig. \ref{Fig2}(c) visualizes the measured weights on two MRRs using ChiL and traditional LUTs. The weights obtained using ChiL, shown as blue points, closely match the target weights, demonstrating the significantly higher programming accuracy of ChiL compared to traditional LUT approaches. In contrast, the measured weights from LUTs, shown as gray points, deviate significantly from the targets. These deviations are primarily caused by thermal crosstalk. For example, when the target weight for MRR 2 is fixed at 0.5 and the target weight for MRR 1 changes from 1 to 0.1, the increasing current on MRR 1 transfers more heat to MRR 2, causing the measured weights on MRR 2 to gradually drift further from 0.5. Meanwhile, the measured weight on MRR 1 also deviates from the target due to the heat from MRR 2. However, this thermal crosstalk issue is effectively mitigated by ChiL. To quantify the programming precision, we define it as bit precision given by:
\begin{equation}
    Bit=log_2(\frac{w_{max}-w_{min}}{RMSE_w})
    \label{eq3}
\end{equation}
where $RMSE_w=\sqrt{<(w-\hat{w})^2>})$, $w$ is the measured weight, $\hat{w}$ is the target weight, $<\cdot>$ denotes average. A record-high bit precision of 9.4-bit is achieved using ChiL, in contrast to the 5.4-bit precision obtained by LUTs. The error distribution using ChiL is plotted in Fig. \ref{Fig2}(d).

We further assess the scalability of ChiL in terms of its precision and convergence speed as the device number increases. We test three weight banks of sizes 2×2, 4×2, and 4×4, corresponding to 4, 8, and 16 weights, respectively. For each weight bank, we test 100 groups of random target weights, with each group tested three times to assess dynamic errors. As the weight bank size increases, thermal crosstalk becomes more pronounced, resulting in the precision using LUT degrades from 5.4-bit to 1.5-bit, as shown in Fig. \ref{Fig2}(e). However, ChiL maintains the bit precision above 9-bit as the weight number increases from 2 to 16, showing it can effectively handle thermal crosstalk in large scale MRR circuits. The error distribution for 16 weights using ChiL is shown as violin plots in Fig. \ref{Fig2}(f). Each curve represents the probability density of weight errors, with the central box denoting the median, interquartile range, and 95\% confidence interval. These results demonstrate that ChiL can be effectively applied to large-scale MRR circuits, despite challenges from thermal crosstalk and device variances.

Our method is well-suited for large-scale PICs due to not only its high precision but also its fast convergence speed. Fig. \ref{Fig2}(g) and Fig. \ref{Fig2}(h) show the iteration history of control variables $I^2$ and weights in the process of programming the 4×4 weight bank. All variables are updated in parallel in one iteration. In Fig. \ref{Fig2}(h), the solid lines represent the measured weights during iteration, and the dashed lines represent the target weights. It can be observed that all weights converge to the target weights in 9 iterations. Fig. \ref{Fig2}(I) shows the averaged RMSE of all weights and the corresponding precision history in 100 experiments of programming the 4×4 MRR weight bank. 9-bit precision is achieved within about 10 iterations. 

\begin{figure}[ht!]
\centering\includegraphics[width=1\linewidth]{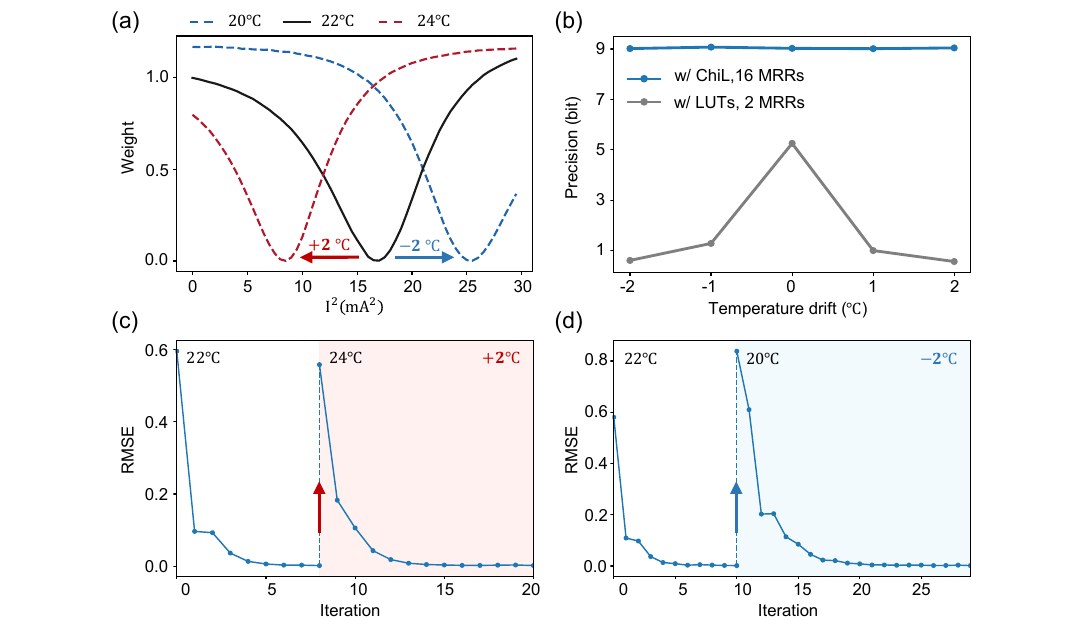}
\caption{
\textbf{Experimental results of temperature robustness test} (a) Change of weight-current tuning curve under temperature drift. (b) Precision of programming the MRR weight bank to 100 groups of target weights under different levels of temperature drift using ChiL and LUTs. (c-d) Weight error correction process of a 4×4 weight bank under positive(c) and negative(d) temperature drift of 2°C
}
\label{Fig3}
\end{figure}

\subsection{Demonstration of Temperature robustness}
The sensitivity of MRRs presents both advantages and challenges. On the one hand, MRRs offer high tuning efficiency; on the other hand, even slight temperature fluctuations can lead to significant weight errors. In Fig. \ref{Fig3}(a), the black solid line shows the weight-current tuning curve of an MRR calibrated at 22°C. A temperature drift of ±2°C causes a dramatic shift, resulting in the weight changing from around 0 to 1 despite the control current remaining constant (see the red and blue dash lines). Traditional methods address this issue by recalibrating all LUTs whenever the ambient temperature changes, adding significant complexity.

In contrast, ChiL is robust to temperature drift, because the model we use to program the MRR circuits, different from LUTs, is irrelevant to temperature. Experimentally, we show that, although our model is calibrated at 22°C, it can accurately program MRR circuits across a wide range of temperatures. We use a thermoelectric cooler  controller (TEC) to simulate different levels of ambient temperature drift. In each temperature, 100 randomly generated target weights are tested for three times. The experiment results show ChiL maintains 9-bit precision on a 4×4 weight bank under both positive and negative temperature drift, while precision achieved with LUTs on a 2×1 weight bank declines from 5.4-bit to less than 1-bit (Fig. \ref{Fig3}(b)).

Moreover, by incorporating the chip within the optimization loop, temperature-induced weight errors can be monitored and corrected in real time. Fig. \ref{Fig3}(c) and Fig. \ref{Fig3}(d) show the recovery process of a 4×4 weight bank under positive and negative temperature drift of 2°C, respectively. The results show that ChiL can rapidly identify a new set of currents to reprogram the weight bank, effectively restoring the weights to their target values within only several iterations. Stronger temperature drift can be further compensated, as long as the target weight can be attained within the feasible region of the control variable (See Supplementary 1 for more details).

\subsection{Photonic eigenvector and eigenvalue solver}
Calculating eigenvectors and eigenvalues is a fundamental problem in linear algebra with numerous applications. Numerical methods, such as power iteration and deflation\cite{golub_matrix_2013}, are commonly used to compute the eigenvectors and eigenvalues of a matrix. The computational complexity of using the power iteration to solve the principal eigenvector and eigenvalue is $O(kN^2)$, where $N$ is the matrix size and $k$ is the number of iterations. For large-scale matrices, solving all eigenvectors and eigenvalues can be time-consuming using digital computers. Analog optical computing offers the potential to accelerate this process, but its accuracy heavily depends on precise control of the photonic system because computing errors at each iteration can accumulate and finally lead to an unacceptable deviation.

\begin{figure}[ht!]
\centering\includegraphics[width=1\linewidth]{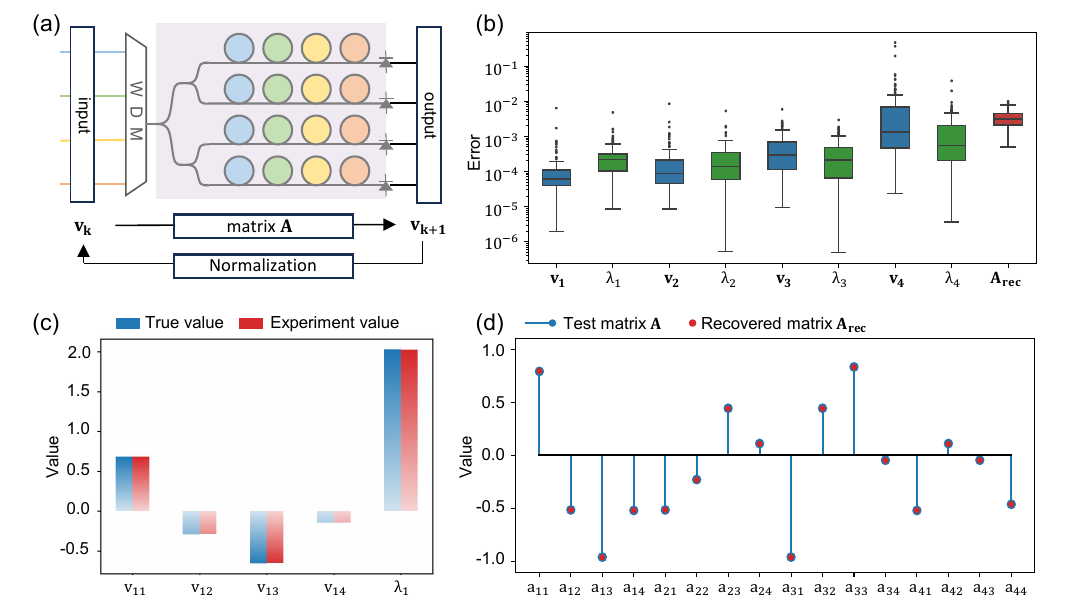}
\caption{
\textbf{Photonic eigenvector and eigenvalue solver using MRR weight bank } (a) Concept of photonic eigenvector and eigenvalue solver using power iteration descrbed in Eq. \ref{eq4}. The Matrix-vector multiplication $\bm{A}\bm{v}_k$ is calculated by the MRR weight bank, where $\bm{v_k}$ will eventually converges to the eigenvector $\bm{v}$ of the matrix. (b) Box plot of eigenvectors ($\bm{v}_i$), eigenvalues ($\lambda_i$), and eigendecomposition ($\bm{A}_{rec}$) error distribution in 100 experiments. (c) Comparison of experimental and theory value of the first eigenvector $\bm{v}_1=[v_{11},v_{12},v_{13},v_{14}]^T$ and the first eigenvalue $\lambda_1$ in one experiment. (d) Comparison of the matrix $\bm{A}_{rec}$ recovered from the eigenvectors and eigenvalues obtained from the weight bank experiment and the original matrix $\bm{A}$. WDM: Wavelength-division multiplexer.
}
\label{Fig4}
\end{figure}

Here we demonstrate the use of high-precision MRR weight bank to solve the eigenvectors and eigenvalues of a matrix through power iteration and deflation, as illustrated in Fig. \ref{Fig4}(a). To solve the eigenvectors and eigenvalues of a matrix $\bm{A}\in\mathbb{R}^{n\times n}$, we first program the matrix onto the MRR weight bank using ChiL. The power iteration process then operates as follows:
\begin{equation}
    \bm{v}_{k+1}=\frac{\bm{A}\bm{v}_k}{\|\bm{A}\bm{v}_k\|_2}
    \label{eq4}
\end{equation}
where $\bm{A}$ is the matrix under evaluation, $\bm{v}_k\in\mathbb{R}^n$ is the vector at the $k$-th iteration, which eventually converges to the eigenvector $\bm{v}$ of the matrix, $\|\cdot\|$ denotes $L^2$ norm. The matrix-vector multiplication $\bm{A}\bm{v}_k$in Eq. \ref{eq4} is calculated by the MRR weight bank. Although optical intensity is inherently positive, real-value matrix-vector multiplication can be performed by linear scaling (See supplementary 1). Once the iteration converges to the eigenvector $\bm{v}$, the corresponding eigenvalue $\lambda$ can be determined by:
\begin{equation}
    \lambda=\frac{\bm{v}^T\bm{A}\bm{v}}{\bm{v}^T\bm{v}}
    \label{eq5}
\end{equation}
We define the error of eigenvector and eigenvalue as:
\begin{equation}
    Error_v=1-\frac{\lvert\bm{v}^T_{exp}\bm{\hat{v}}\rvert}{\|\bm{v}_{exp}\|_2\|\bm{\hat{v}}\|_2}
    \label{eq6}
\end{equation}
\begin{equation}
    Error_{\lambda}=\lvert\lambda_{exp}-\hat{\lambda}\rvert
    \label{eq7}
\end{equation}
where $\bm{v}_{exp}$ and $\lambda_{exp}$ are eigenvector and eigenvalue obtained in experiment; $\bm{\hat{v}}$ and $\hat{\lambda}$ are the theoretical eigenvector and eigenvalue calculated using a digital computer. To experimentally evaluate the accuracy of our eigenvector and eigenvalue solver, we test it on 100 randomly generated matrices. The average error for the first eigenvector and eigenvalue across these matrices is $1.8\times10^{-4}$ and $3.5\times10^{-4}$ , respectively. Benefiting from ChiL, the errors are 1000 times smaller than those reported in a previous work\cite{liao_matrix_2022} that uses a PNN for eigenvalue prediction. Fig. \ref{Fig4}(c) compares the first eigenvector and eigenvalue solved by the MRR weight bank with those computed by a digital computer, demonstrating the results from the weight bank closely matches the theoretical values. 

We further solve all eigenvectors and eigenvalues of these matrices by combining deflation and power iteration. The deflection is expressed as:$\bm{A}=\bm{A}-\lambda\bm{v}\bm{v}^T$, where $\lambda$ and $\bm{v}$ are eigenvalue and eigenvector previously solved. Then power iteration based on Eq. \ref{eq4} can be executed on the new matrix $A$ to solve the next eigenvalue and eigenvector. Symmetric matrices are chosen in our experiments because an n×n symmetric matrix is guaranteed to have n eigenvalues and eigenvectors. Box plots in Fig. \ref{Fig4}(b) show the error distribution of eigenvectors and eigenvalues obtained from the weight bank. To quantify the error in solving all eigenvectors and eigenvalues of a matrix, we introduce the eigendecomposition error, which is defined as difference between the matrix reconstructed from the eigenvectors and eigenvalues obtained from the weight bank experiment and the original matrix. The reconstructed matrix is expressed as: $\bm{A}_{rec}=\bm{V}_{exp}\bm{\Lambda}_{exp}\bm{V}_{exp}^T$, where $\bm{A}_{rec}\in\mathbb{S}^{n\times n}$, $\bm{V}=[\bm{v_1},\bm{\cdots},\bm{v_m}]$, $\bm{\Lambda}=Diag(\lambda_1,\cdots,\lambda_n)$. $\bm{v}_i$ and $\lambda_i$ denote the $i$-th eigenvector and eigenvalue, respectively. The eigendecomposition error is then defined as:
\begin{equation}
    Error_{Dec}=\sqrt{<(a_{rec,ij}-a_{ij})>}
    \label{eq8}
\end{equation}
where $a_{rec,ij}$ and $a_{ij}$ are the elements in the matrix $\bm{A}_{rec}$ and $\bm{A}$, respectively. The average eigendecomposition error evaluated using 100 randomly generated matrices is $3.5\times10^{-3}$. Fig. \ref{Fig4}(d) illustrates a sample showing the matrix elements of the reconstructed matrix (red dots) closely match those in the original matrix (blue dots). The results demonstrate the weight banks programmed by ChiL can solve eigenvectors and eigenvalues with high accuracy.

\begin{figure}[ht!]
\centering\includegraphics[width=1\linewidth]{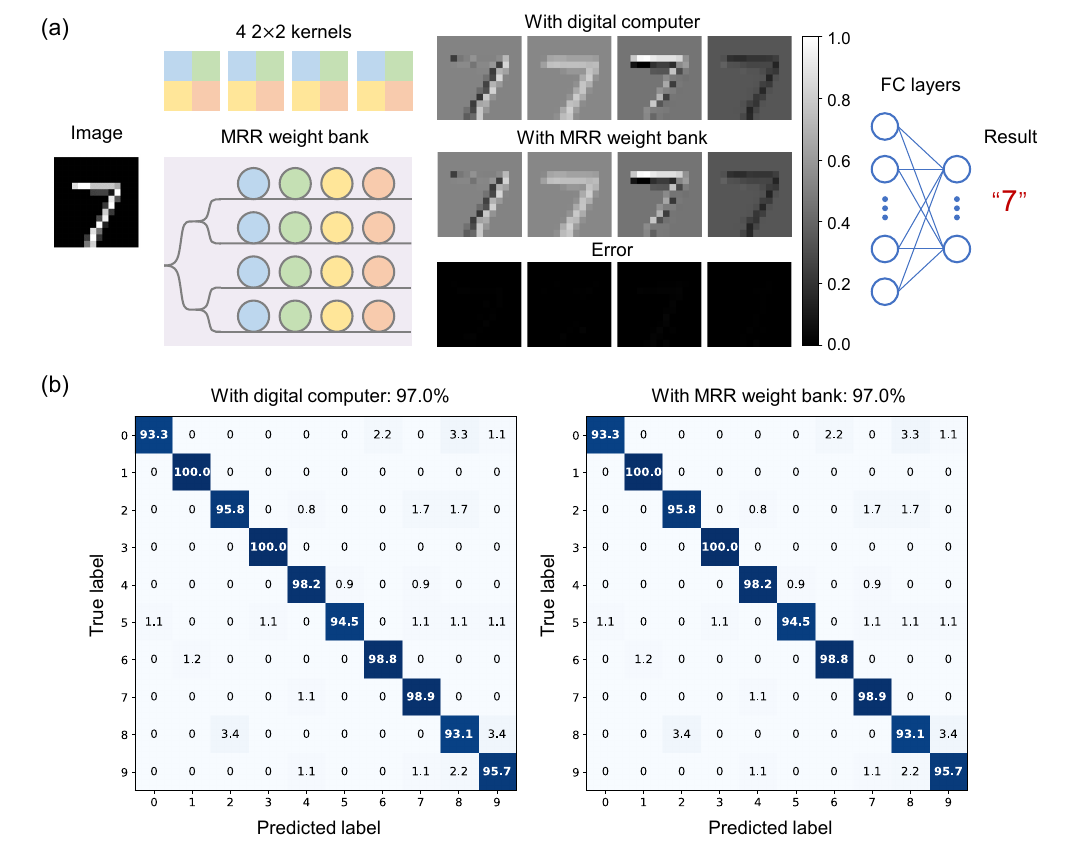}
\caption{
\textbf{PNN for MNIST handwriting digits classification} (a) Concept of PNN. The 4×4 MRR weight bank act as four 2×2 kernels in the convolutional layer. The convoluted outputs obtained from the weight bank closely match those from the digital computer. (b) Confusion matrix obtained from the PNN and digital NN. FC layers: fully-connected layers.
}
\label{Fig5}
\end{figure}

\subsection{Photonic neural networks}
Artificial neural networks have made great success in natural language processing, computer vision, scientific discovery, and many other fields in recent years\cite{lecun_deep_2015}. However, the increasing scale of NNs lays great computing power and energy consumption challenges to traditional digital electronic computing for the end of Moore’s law\cite{mehonic_brain-inspired_2022}. PNNs, featuring large bandwidth, low latency, and low power consumption, present as a promising solution\cite{zhou_photonic_2022,fu_optical_2024}. However, the difficulties of precisely programming photonic hardware always make PNNs less accurate than their digital counterparts\cite{xu_control-free_2024}. While strategies like optical pruning\cite{xu_control-free_2024} and noise-aware training\cite{wu_harnessing_2022} can help narrow this gap, they require more trainable parameters and increased training costs to reach the robust parameter space. Here, we demonstrate that the high computing accuracy enabled by ChiL allows for the direct implementation of a digitally trained NN on an MRR weight bank without loss in accuracy--not only the classification accuracy but also the confusion matrix is identical to those achieved on a digital computer.

We demonstrate the PNN on the 10-class MNIST handwriting digits classification task, as shown in Fig. \ref{Fig5}(a). A convolutional neural network is designed and trained with standard backpropagation on a computer, achieving a test accuracy of 97.0\% (see Supplementary 1 for more details). Then we directly program the convolution layer, which employs four 2×2 kernels, on the 4×4 MRR weight bank using ChiL. Each image is first down sampled to 14×14 pixels, then repeated and reshaped into 196 groups of 4×1 vectors and modulated on four wavelengths as the input signals. An identity matrix follows to monitor the current weights. Once the monitored weights deviate from the target weights due to dynamic noise, the control parameters will update to correct them. As shown in Fig. 5a, the convoluted outputs obtained from the weight bank closely match those from the digital computer, with no observed pixel error. We achieve a classification accuracy of 97.0\% with the PNN, which is identical to that of the digital NN. Remarkably, the confusion matrix obtained from the PNN is also identical to that of the digital NN, without extra classification error, as shown in Fig. \ref{Fig5}(b). These results demonstrate that the PNNs controlled by ChiL can achieve the numerical precision match with digital computers.

\begin{table}[ht!]
\centering
\caption{\bf Comparison of this work with previous works}
\begin{tabular}{cccc}
\hline
Method & Precision × Scale & \makecell[cc]{Calibration\\complexity} & \makecell[cc]{Temperature\\robustness} \\
\hline
N-doped heater monitoring\cite{huang_demonstration_2020} & 7.2-bit × 4 & $O(kN)^\textit{a}$ & N/A\\
Dithering control\cite{zhang_silicon_2022} & 9.0-bit × 2 & $O(kN^3)^\textit{b}$ & N/A \\
Dual-wavelength synchronization\cite{cheng_self-calibrating_2023} & 7.0-bit × 9 & $O(kN)^\textit{a}$ & 0.5℃\\
Single-monitor calibration\cite{liu_single-monitor_2024} & 4.5-bit$^\textit{c}$ × 16 & $O(kN)^\textit{a}$ & 0.2℃ \\
Gradient-decent control\cite{bai_microcomb-based_2023} & 9.0-bit × 4 & $O(kN)^\textit{a}$ & N/A \\
ChiL (This work) & 9.1-bit × 16 & $O(1)$ & 4℃\\
\hline
\end{tabular}
\label{table1}

$^\textit{a}$Complexity of building LUTs for each MRR, without considering thermal crosstalk.\\
$^\textit{b}$Complexity of matrix inversion computation needed during calibration.\\
$^\textit{c}$Calculated from average weight error provided in the paper. 

\end{table}

\section{Discussion and conclusion}
Precise programming of imperfect photonic systems with simple, fast, and low-cost methods is crucial for the practical deployment of large-scale PICs. With ChiL, we significantly reduce the complexity in calibration, which typically scales exponentially with the number of devices, to a single model that requires only one measurement on one device. This model can be applied universally—not only across multiple devices on the same chip but also across different chips, even accounting for fabrication variations. At the same time, ChiL eliminates the need for additional monitor devices used in previous works, such as source meters\cite{tait_feedback_2018,jayatilleka_photoconductive_2019,huang_demonstration_2020}, dithering signal generators\cite{zhang_silicon_2022}, or extra lasers and photodetectors\cite{cheng_self-calibrating_2023,liu_single-monitor_2024}. This not only reduces control costs but also prevents errors introduced by the monitoring devices.

We compare ChiL with recent approaches in Table 1. To the best of our knowledge, our approach achieves the largest scale, highest precision, and most robust MRR programming. The programing precision can be further enhanced by using high-quality lasers and packaging the chip with optical fiber I/O to minimize input power and polarization fluctuations. 

Our high-precision, large-scale MRR circuit elevates PIC performance across multiple applications. For numerical calculations, we successfully compute eigenvectors and eigenvalues of matrices with 1000 times of error reduction. For artificial intelligence, we demonstrate a PNN with no accuracy loss compared to a digital computer. We show that ChiL bridges the gap of computing precision between analog photonic computing and digital electronic computing. Moreover, our method has the potential to enhance other PIC applications, including optical interconnects, optical switches, microwave and optical signal processing, and many more.

\begin{backmatter}

\bmsection{Funding}
This work was supported RGC ECS 24203724, NSFC 62405258, ITF ITS/237/22, ITS/226/21FP, RGC YCRF C1002-22Y, RNE-p4-22 of the Shun Hing Institute of Advanced Engineering, NSFC/RGC Joint Research Scheme N CUHK444/22 and CUHK Direct Grant 170257018, 4055143.

\bmsection{Disclosures}
The authors declare no conflict of interest.

\bmsection{Data availability}
The data and code that support this study are available from the corresponding authors upon reasonable request.

\end{backmatter}

\bibliography{MRR_Precision}

\newpage
\appendix
\title{High-precision programming of large-scale ring resonator circuits with minimal pre-calibration: supplemental document}
\author{} 

\renewcommand{\thesection}{\arabic{section}}
\section{Chip design, fabrication and package}
The photonic chip is fabricated on the silicon-on-insulator (SOI) platform with a 220-nm-thick silicon device layer and 2.2-um-thick oxide cladding in Applied Nanotools (ANT). The MRR weight bank consists of 16 MRRs arranged as 4 rows. Each row has 4 MRRs with radii of 20.00um, 20.01um,20.02um, and 20.03um, coupled with 2 bus waveguides in an add/drop configuration. The gap between MRRs and bus waveguides is 0.1 um. MRRs’ Q factor is about 10,000. Light inputs and outputs through TE mode grating couplers provided by ANT’s PDK, with around -7dB coupling loss each side. Titanium-tungsten (TiW) alloy heater is built on top of each MRR for thermal tuning. The heater’s resistance is around 380 ohm. Each heater is connected to TiW/Al routings, which are wire-bonded to pads on a print circuit board (PCB).

\section{Experimental setup}
We use four tunable lasers (KG-TLS) as light sources with working wavelengths of 1542.2nm, 1542.9nm, 1543.6nm, 1544.3nm and output power of 10.00dBm. Four variable optical attenuators (OE-VOA) are then used to generate input signals of the 4 channels. After going through four polarization controllers, they are combined together by three 5:5 couplers, and then coupled into the photonic chip through a 16-channel fiber array without an optical package. The four output ports are detected by a 4-channel optical power meter (Keysight N7744C). A 64-channel current source (Silicon Extreme MSVS6400-A) is used to tune the VOAs and microheaters on MRRs.

\section{Measuring MRRs’ weight from system’s output}
The transfer function of an MRR weight bank can be described as a weight matrix $\bm{W}\in\mathbb{R}^{m\times n}$. Take $m=n=4$ as example. If we set the input vector $\bm{x}=[1,0,0,0]^T$, at the output we get the first column of the matrix $\bm{y}=[w_{11},w_{12},w_{13},w_{14}]$. Then we change the input vector to be  $[0,1,0,0]^T$,  $[0,0,1,0]^T$, and $[0,0,0,1]^T$sequentially to measure the other columns of weights. This procedure is described as $\bm{Y}=\bm{WI}=\bm{W}$, where $\bm{I}$ is the identity matrix.

\section{Build the model for MRRs}
The model consists of two key parameters of MRRs: (1) tuning efficiency and (2) gradient range.  The tuning efficiency is used to apply the initial current on each MRR, and the gradient range is used to constrain the estimated gradient in the optimization loop. We characterize one MRR to get these parameters.

\begin{figure}[ht!]
\centering\includegraphics[width=1\linewidth]{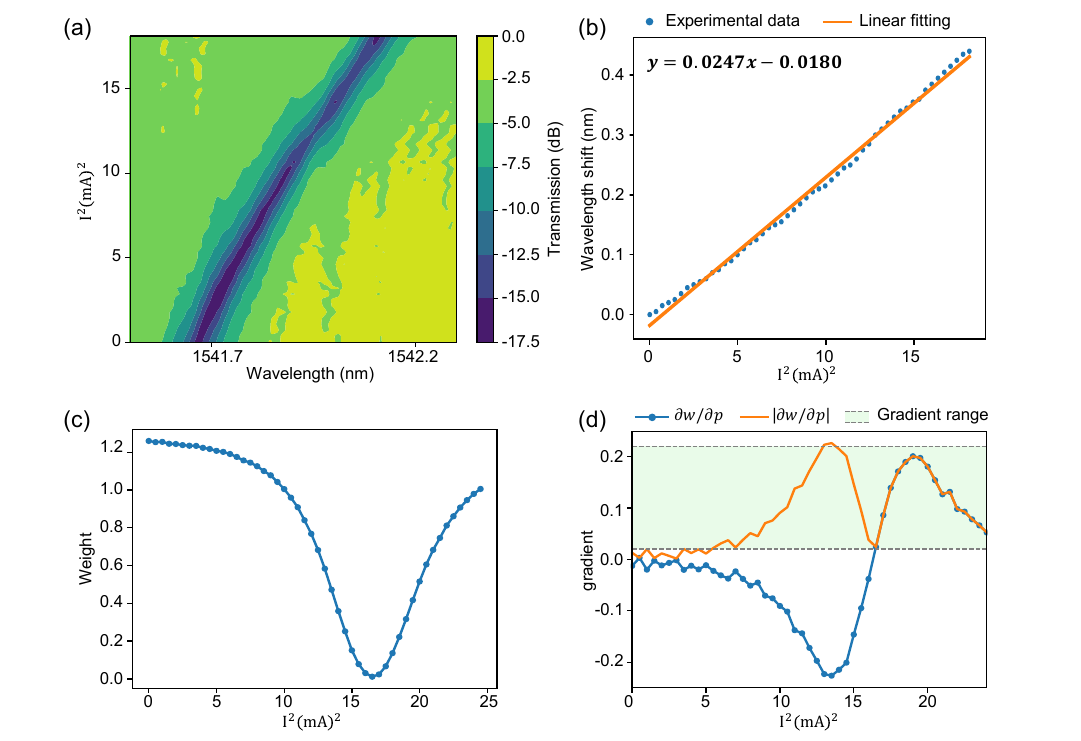}
\caption{
\textbf{Build the model for MRRs} (a) An MRR’s spectrum when tuning current. (b) An MRR’s resonant wavelength shift when tuning current. (c) weight-current tuning curve of the MRR. (d) gradient range of the MRR. 
}
\label{Fig S1}
\end{figure}

To get the tuning efficiency, first, we use ASE noise as a light source and an optical spectrum analyzer (OSA) to measure the spectrum of the MRR. Then we tune the current that is applied to the microheater on the MRR. Fig. \ref{Fig S1}(a) shows the spectrum change while tuning the current. The resonant wavelength of the MRR redshifts when the current increases. We use linear fitting to describe the relationship between resonant wavelength shift and square of current, as shown in Fig. \ref{Fig S1}(b). We get the tuning efficiency of the MRR is $\gamma=0.0247nm/mA^2$. 

To get the gradient range, we use a laser with a working wavelength at the right of the MRR’s resonant wavelength. Then we tune the current that applied to the the MRR while using an optical power meter to monitor its weight. The weight-current tuning cure is shown in Fig. \ref{Fig S1}(c). Then we do a finite difference to estimate its gradient: $\frac{\partial w_k}{\partial p_k} = \frac{\partial w_{k+1}-\partial w_{k}}{p_{k+1}-p_{k}}$ , where $p=I^2$ is the control variable. The gradient-current curve is shown in Fig. \ref{Fig S1}(d). We set the absolute gradient range to be $[G_min,G_max ]=[0.02,0.22] mA^{-2}$ 

\section{Process of chip initialization}
Due to the fabrication variance, MRRs’ spectra may alias with each other. We set a group of currents to decouple these spectra and align the resonant wavelength of MRRs in the same column, which is called chip initialization. First, we calculate the required currents by$ I=\sqrt{(\bm{\lambda_t}-\bm{\lambda_0})/\gamma}$ , where $\bm{\lambda_t}$ are the target resonant wavelengths, $\bm{\lambda_0}$ are the resonant wavelength at zero currents, and $\gamma$ is the tuning efficiency from the model. After applying these currents, the resonant wavelengths usually deviate from the targets because of the inaccurate model and thermal crosstalk. We can further tune the currents to align the resonant wavelengths by the following steps: (1) Set the lasers’ wavelengths to the target resonant wavelengths. (2) Define the target matrix $\bm{W}=\bm{0}$ . (3) Execute ChiL. After that, all resonant wavelengths are aligned well, as shown in Fig. 2b in the main text. Now we get a set of currents for chip initialization. The above procedure to find them just needs to be done once for a chip. 

\section{Programing MRRs with LUTs}
\begin{figure}[ht!]
\centering\includegraphics[width=1\linewidth]{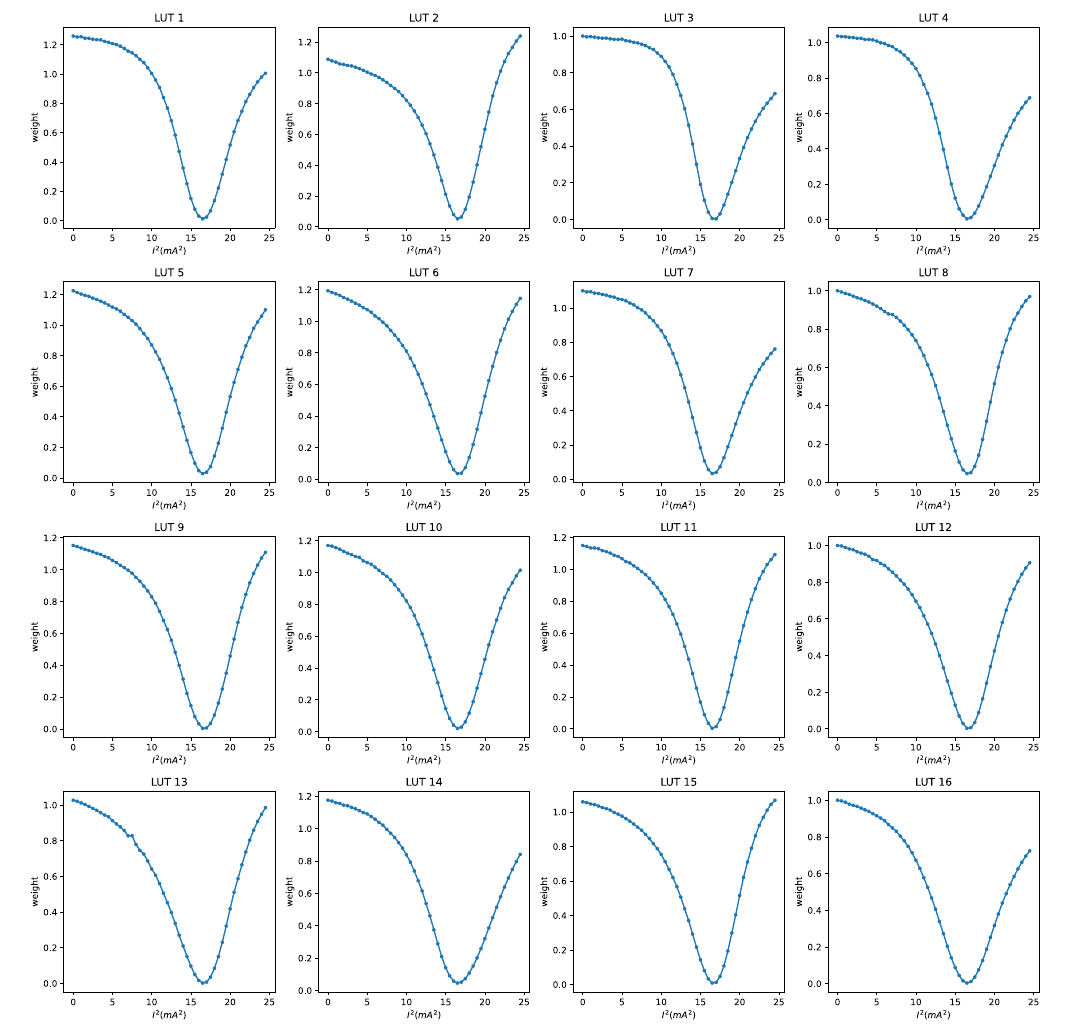}
\caption{\textbf{LUTs of 16 MRRs}}
\label{Fig S2}
\end{figure}

The way we build the LUTs is the same as previous work\cite{huang_demonstration_2020}. We sequentially sweep the current of each MRR and measure its transmission, while keeping currents on other MRRs to be 0. The total number of measurement points is $kN$, where $k$ is the number of points in each LUT, and $N$ is the number of MRR. Thermal crosstalk is not calibrated here.  Fig. \ref{Fig S2} shows the weight- current LUT for each MRR, where $k = 50$ and $N = 16$. When given commanded weights that are not included in these points, we use linear interpolation to determine the required current.

\section{Programing MRR circuits under temperature drift}
MRR’s spectrum red-shifts when temperature increases and blue-shifts when temperature decreases. A ±2°C temperature drift is demonstrated in this work, but it is not the limit of our method. The precision of programming MRR circuits under temperature drift is limited by the feasible region of the control variables. In this work, the control variable is the square of currents. First, it must be larger than 0, meaning that applying current on an MRR can only red-shift its spectrum. If the temperature increases so much that the MRR’s resonant wavelength shifts to the right of the laser’s wavelength, the weight 0 can’t be attained anymore. Second, currents must be smaller than the maximum tolerance of electrical design. If the temperature decreases too much, the maximum current may not be able to compensate for the severe blue-shift of the spectrum. 

\begin{figure}[ht!]
\centering\includegraphics[width=1\linewidth]{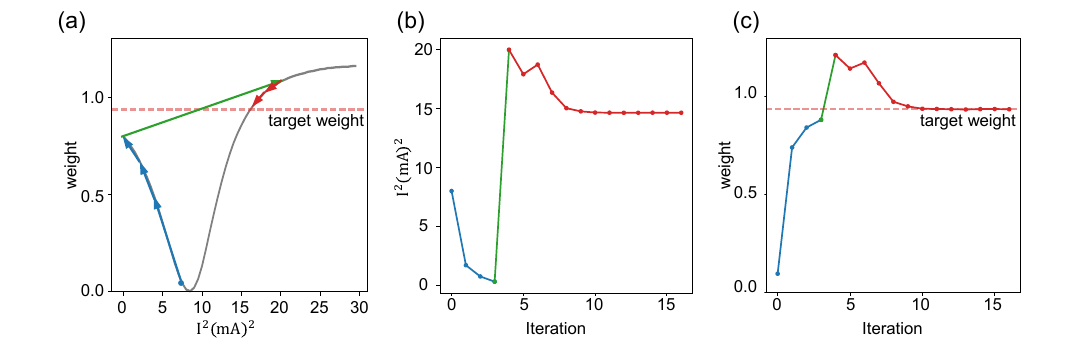}
\caption{
\textbf{An example of programming MRR when temperature increases for 2 ℃} (a) Schematic diagram of the search point moving on the tuning curve. (b-c) Current and weight change during the iteration when programing a target weight of 0.9367.
}
\label{Fig S3}
\end{figure}

Traditionally, the left side of the tuning curve is utilized to achieve the target weight due to the lower power consumption for the same weight compared to the right side. However, a special case to note when dealing with increased temperatures is that, sometimes even when the current is reduced to 0 A, the resulting weight may still be smaller than the target value. To address this issue, we use the following strategy: if the measured weight remains below the target even at zero current, we apply a large current to shift the operating point from the left side to the right side of the tuning curve, which ensures the convergence to the target weight. Fig. \ref{Fig S3} shows an example of programming an MRR when the temperature increases by 2℃. The target weight is 0.9367, which can’t be attained on the left of the tuning curve, as shown in Fig. \ref{Fig S3} (a). There are three iteration phases, shown as blue, green, and red lines separately. In phase one, the measured weight is smaller than the target weight and the gradient is negative, where the current decreases to reduce weight error. However, the measured weight is still smaller than the target weight when the current is 0. Then in phase two, we set $p=I^2=20$, which brings the searching point to the right of the tuning curve and makes the measured weight larger than the target weight. The sign of gradient is also reversed. Finally, in phase three, the current decreases to make the measured weight closer to the target weight. We get a measured weight of 0.9365.

\section{Real-value matrix-vector multiplication}
We use linear scaling to transfer positive transmission (between 0 and 1) to the real-value weight we want. This method is similar to Ref. \cite{feldmann_parallel_2021}. Assumed that transmission of an MRR is $t\in[t_{min},t_{max}]$, and we want to get the weight: $w\in[-1,1]$. We can define the weight as:
\begin{equation}
    w=\frac{1-(-1)}{t_{max}-t_{min}}t+\frac{(-1)t_{max}-1t_{min}}{t_{max}-t_{min}}=kx+b
\end{equation}
where $k=\frac{2}{t_{max}-t_{min}}$ and $b=-\frac{t_{max}+t_{min}}{t_{max}-t_{min}}$ is constant.Given a positive input vector $\bm{x}$ and a real-value target weight matrix $\bm{W}$, we first calculate the transmission matrix by: $\bm{T}=(\bm{W}-b)/k$, then program it on the MRR weight bank. After getting theoutput $\bm{y}=\bm{Tx}$, we stretch it by $\bm{\hat{y}}=k\bm{y}+b\bm{x}=\bm{Wx}$. For a real-value input vector $\bm{x}$, we decompose it to positive part and negative part by: $\bm{x}=\bm{x}^+ + \bm{x}^-$,$\bm{x}^+\geq0$,$\bm{x}\leq0$. Then we do matrix-value multiplication on both these two parts with above method and subscribe the results by: $\bm{\hat{y}}=\bm{Wx}=\bm{Wx}^+-\bm{W}(-\bm{x}^-)$. Such that real-value matrix-vector multiplication is supported in our scheme.

\section{Design and Training of the NN for MNIST classification}
The NN for MNIST handwritten digits classifications consists of one convolutional layer and two fully connected layers. The convolutional layer employs four 2×2 kernels, while the fully connected layers have sizes of 676×100 and 100×10, respectively. ReLU nonlinear activation is applied between each layer. The input images are down sampled to 14×14 pixels. We construct and train the NN using PyTorch. The cross-entropy loss function is employed to evaluate its performance. The Adam optimizer is utilized for parameter updates with a learning rate set to 0.01. We use 50000 samples, with a batch size of 128, for training and 1000 samples for testing. After training for 10 epochs, we get an accuracy of 97.0\%.
\end{document}